\documentclass[a4paper,nocompress]{spie}

\usepackage{amsmath,amsfonts,amssymb}
\usepackage{graphicx}
\usepackage[colorlinks=true, allcolors=blue]{hyperref}
\usepackage{txfonts}
\usepackage{adjustbox}
\usepackage{subfigure}
\usepackage{xspace}
\usepackage{amsmath,amssymb}

\newcommand{\mic}{\ensuremath{\muup\mathrm{m}}\xspace}
\newcommand{\as}{\hbox{$^{\prime\prime}$}\xspace}

\title{Connecting SPHERE and CRIRES+ for the characterisation of young exoplanets at high spectral resolution: status update of VLT/HiRISE}

\author[a]{A.~Vigan}
\author[a]{M.~Lopez}
\author[a]{M.~El Morsy}
\author[a,i,j]{E.~Muslimov}
\author[a]{A.~Viret}
\author[b]{G.~Zins}
\author[c]{G.~Murray}
\author[a]{A.~Costille}
\author[a,k]{G.~P.~P.~L.~Otten}
\author[b,d]{U.~Seemann}
\author[d]{H.~Anwand-Heerwart}
\author[a]{K.~Dohlen}
\author[a]{P.~Blanchard}
\author[a]{J.~Garcia}
\author[a]{Y.~Charles}
\author[a]{N.~Tchoubaklian}
\author[a]{T.~Ely}
\author[e]{M.~Phillips}
\author[b]{J.~Paufique}
\author[a]{J.-L.~Beuzit}
\author[a]{M.~Houllé}
\author[a]{J.~Costes}
\author[a,h]{R.~Pourcelot}
\author[e]{I.~Baraffe}
\author[b]{R.~Dorn}
\author[a]{M.~Jaquet}
\author[b]{M.~Kasper}
\author[d]{A.~Reiners}
\author[f]{A.~Smette}
\author[f]{L.~Blanco}
\author[f]{L.~Pallanca}
\author[g]{A.~Carlotti}
\author[a]{\'E.~Choquet}
\author[g]{D.~Mouillet}
\author[h]{M.~N'Diaye}

\affil[a]{Aix Marseille Univ, CNRS, CNES, LAM, Marseille, France}
\affil[b]{European Southern Observatory (ESO), Karl-Schwarzschild-Str. 2, 85748 Garching, Germany}
\affil[c]{Center for Advanced Instrumentation, Durham University, Durham, DH1 3LE, United Kindgom}
\affil[d]{Institute for Astrophysics, Georg-August University, Friedrich-Hund-Platz 1, 37077 Göttingen, Germany}
\affil[e]{School of Physics and Astronomy, University of Exeter, Exeter EX4 4QL, UK}
\affil[f]{European Southern Observatory (ESO), Alonso de C\'ordova 3107, Vitacura, Casilla 19001, Santiago, Chile}
\affil[g]{Univ. Grenoble Alpes, CNRS, IPAG, 38000 Grenoble, France}
\affil[h]{Universit\'e C\^ote d'Azur, Observatoire de la C\^ote d'Azur, CNRS, Laboratoire Lagrange, Bd de l'Observatoire, CS 34229, 06304, Nice Cedex 4, France}
\affil[i]{NOVA Optical IR Instrumentation Group at ASTRON Oude Hoogeveensedijk 4, 7991 PD Dwingeloo, The Netherlands}
\affil[j]{Kazan National Research Technical University named after A.N. Tupolev KAI, 10 K. Marx, Kazan, Russia, 420111}
\affil[k]{Academia Sinica, Institute of Astronomy and Astrophysics, 11F Astronomy-Mathematics Building, NTU/AS campus, No. 1, Section 4, Roosevelt Rd., Taipei 10617, Taiwan}

\authorinfo{Send correspondence to A. Vigan (\href{mailto:arthur.vigan@lam.fr}{arthur.vigan@lam.fr}).}

\begin{document} 
\maketitle

\begin{abstract}
New generation exoplanet imagers on large ground-based telescopes are highly optimised for the detection of young giant exoplanets in the near-infrared, but they are intrinsically limited for their characterisation by the low spectral resolution of their integral field spectrographs ($R<100$). High-dispersion spectroscopy at $R \gg 10^4$ would be a powerful tool for the characterisation of these planets, but there is currently no high-resolution spectrograph with extreme adaptive optics and coronagraphy that would enable such characterisation. With project HiRISE we propose to use fiber coupling to combine the capabilities of two flagship instruments at the Very Large Telescope in Chile: the exoplanet imager SPHERE and the high-resolution spectrograph CRIRES+. The coupling will be implemented at the telescope in early 2023. We provide a general overview of the implementation of HiRISE, of its assembly, integration and testing (AIT) phase in Europe, and a brief assessment of its expected performance based on the final hardware.
\end{abstract}


\section{INTRODUCTION}
\label{sec:intro}

Direct imaging allows to spatially separate and directly measure radiation from an exoplanet, which enables the spectral analysis of its atmosphere with minimized impact from the host star. Direct imagers such as SPHERE\cite{Beuzit:2019}, GPI\cite{Macintosh:2014}, and SCExAO\cite{Jovanovic:2015} are designed to find and detect young planets around nearby stars\cite{Macintosh:2015,Chauvin:2017a,Keppler:2018}, which can be used to study the population of young giant exoplanets through large direct imaging surveys\cite{Vigan:2012,Galicher:2016,Vigan:2017,Vigan:2021}. However, current instruments are limited for the characterization of companions by their spectral resolution of $R=\lambda/\Delta\lambda=400$ at most\cite{Vigan:2008}.

High-dispersion spectrographs have been used to detect the thermal radiation of various giant exoplanets by relying on the large number of molecular lines detectable in the atmospheres of these planets, which can only be accessed at spectral resolutions above a few 10\,000s. Important results have been obtained on both transiting and non-transiting planets\cite{Snellen:2010,Brogi:2012,Birkby:2013}, as well as on spatially resolved low-mass companions\cite{Snellen:2014,Schwarz:2016,Hoeijmakers:2018}. For the latter, the most well-known case is that of the directly-imaged exoplanet $\beta$\,Pictoris\,b\cite{Snellen:2014} using the CRIRES instrument\cite{Kaeufl:2004} on the ESO Very Large Telescope (VLT). This method has since provided more atmospheric detections and rotation speeds of young directly imaged companions\cite{Schwarz:2016,Hoeijmakers:2018,Bryan:2018}, providing unique insight into the properties of this population of objects\cite{Bryan:2018}.

Measurements on young companions have been obtained with low-order adaptive optics systems like MACAO\cite{Arsenault:2003} for the $\beta$~Pictoris\,b result. With such systems, the atmospheric turbulence correction  concentrates a moderate fraction of the energy in the PSF core (typically 50--60\% in the $K$ band\cite{Paufique:2006}), but the level of the uncorrected halo is high. This means that the strongest limiting factor of the signal-to-noise ratio (S/N) of the planet signal close to the star remains the noise contributed by the uncorrected stellar halo. To significantly decrease the halo, it is necessary to rely on high-order adaptive optics known as extreme adaptive optics (ExAO), which can provide diffraction-limited images in the near-infrared (NIR) and therefore decrease the level of stellar halo (and noise) at the location of the planet\cite{Kawahara:2014a,Snellen:2015,Wang:2017,Mawet:2017,Vigan:2018}. Diffraction-suppressing coronagraphs can also be used to further improve the contrast and decrease the level of stellar residuals at the location of the planet.

Until recently, no instrument dedicated to high-contrast imaging was equipped with means of characterizing detected companions at very high spectral resolutions. Conversely, none of the existing high-resolution spectrographs in the near infrared is equipped with either an ExAO system or with a coronagraph, which would facilitate the characterization of known companions. In recent years, several projects have proposed combining ExAO systems with existing mid- or high-resolution spectrographs using single-mode fibers (SMF) either in the NIR, for example between NIRC2 and NIRSPEC at Keck (KPIC)\cite{Mawet:2016,Mawet:2017} or between SCExAO and IRD at Subaru (REACH)\cite{Kawahara:2014,Kawahara:2014a,Kotani:2018}, or in the visible, for example between SPHERE and ESPRESSO at the VLT\cite{Lovis:2017}. For the NIR, the VLT offers a unique opportunity to achieve a similar feat by coupling the high-contrast imager SPHERE\cite{Beuzit:2019} with the recently upgraded high-resolution spectrograph CRIRES+\cite{Dorn:2016}, which are now both available at the VLT/UT3. The coupling between these two flagship instruments has been proposed as the High-Resolution Imaging and Spectroscopy of Exoplanets project (HiRISE)\cite{Vigan:2018,Otten:2021}.

SPHERE offers a unique ExAO system (SAXO)\cite{Fusco:2006} that has demonstrated exquisite performance on-sky\cite{Sauvage:2014,Petit:2014,Milli:2017} and efficient coronagraphs\cite{Carbillet:2011,Guerri:2011}; its infrared arm covers the $Y$, $J$, $H$, and $K_s$ bands. CRIRES+ is a refurbished and upgraded version of the CRIRES spectrograph\cite{Kaeufl:2004}, operating in the $Y$, $J$, $H$, $K$, $L$, and $M$ bands (0.9-5.3\,\mic) at $R=100\,000$ resolving power with a 0.2$''$-slit. It features three Hawaii-2RG 2k$\times$2k detectors and a cross-disperser to cover more than 50\% of the $H$ band in a single observation. The large overlap in terms of spectral coverage between SPHERE and CRIRES+ is a key advantage in particular in the $H$ and $K$ bands, which contain strong molecular features for CO, CH$_4$, H$_2$O, or NH$_3$. The high angular resolution and high-contrast capabilities of SPHERE at these wavelengths, combined with the high spectral resolution of CRIRES+, will provide a unique system capable of characterizing known directly imaged companions.

In this paper we present the final proposed implementation of the system in Sect.~\ref{sec:implementation}, an assessment of the expected performance based on the final hardware in Sect.~\ref{sec:expected_perf}, and finally the current status of the project and its schedule in Sect.~\ref{sec:project_status}.

\section{IMPLEMENTATION}
\label{sec:implementation}

\subsection{Overview}
\label{sec:implementation_overview}

\begin{figure}
  \centering
  \includegraphics[width=0.8\textwidth]{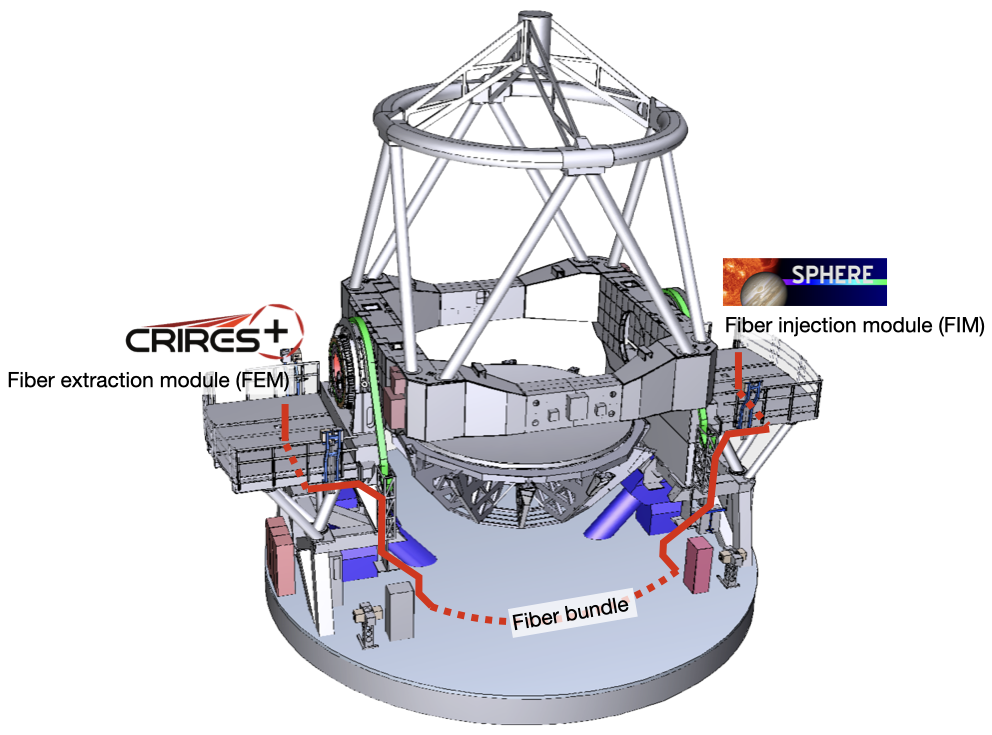}
  \caption{Overview of the implementation at the telescope. SPHERE and CRIRES+ are both located on the VLT Unit Telescope 3 (UT3), on Nasmyth platforms A and B respectively. The bundle cannot be routed along structure that supports the primary mirror due to the altitude motion during science observations. The favored option is to route the bundle along the Nasmyth structures and beneath the floor. The route depicted here is not exactly the final one that has been studied and proposed by ESO.}
  \label{fig:implementation_overview}
\end{figure}

The HiRISE coupling will be implemented on the VLT-UT3 between SPHERE and CRIRES+, which are located on Nasmyth platforms A and B, respectively. This is illustrated in Fig.~\ref{fig:implementation_overview}. The system consists in three main sub-systems:

\begin{enumerate}
    \item the fiber injection module (FIM) implemented inside of SPHERE. The role of the FIM is to pick-up the SPHERE scientific beam and inject the signal of a known exoplanet in a science fiber. The FIM will be described in Sect.~\ref{sec:implementation_fim};
    \item the fiber bundle that connects SPHERE and CRIRES+ around the telescope. The bundle transmits the planetary and stellar signals up to the entrance of the spectrograph. The bundle will be described in Sect.~\ref{sec:implementation_fb};
    \item and the fiber extraction module (FEM) implemented inside of CRIRES+. The role of the FEM is to reimage the output of the bundle's fibers in the VLT focal plane with the appropriate focal ratio. The FEM will be described in Sect.~\ref{sec:implementation_fem};
\end{enumerate}

In addition to these main sub-systems, HiRISE also includes a dedicated electronics cabinet located on the SPHERE Nasmyth platform, and a dedicated control software developed following the VLT Software standards. These additional components will be described in Sect.~\ref{sec:implementation_electronics}.

\subsection{Fiber injection module}
\label{sec:implementation_fim}

\begin{figure}
  \centering
  \includegraphics[width=0.7\textwidth]{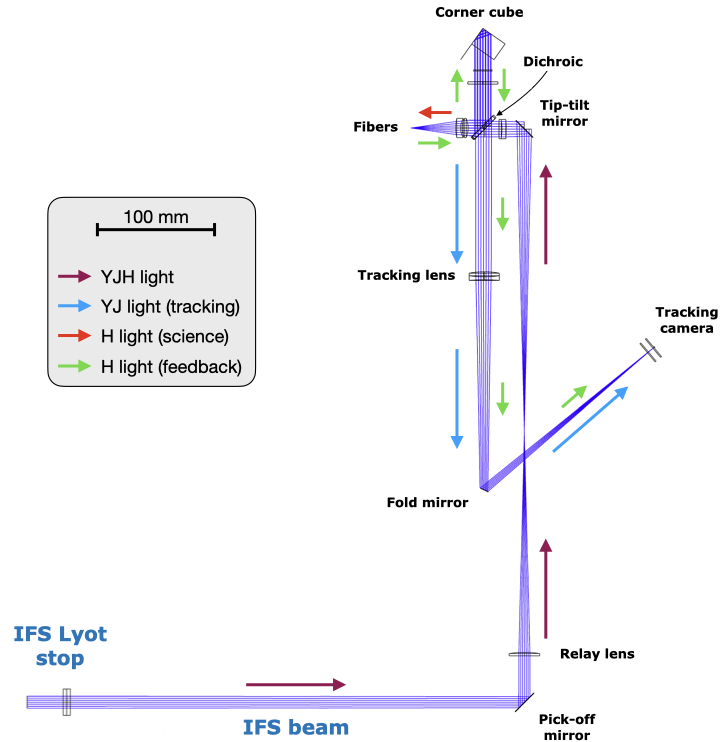}
  \caption{Optical design and photon sharing of the FIM located in SPHERE. The science channel transmits only the $H$ band for the injection into the science fiber. The tracking channel transmits the $Y$ and $J$ bands for acquisition and monitoring with the tracking camera. All the optics are coated with appropriate AR-coatings. Figure borrowed from [\citenum{ElMorsy:2022}].}
  \label{fig:fim_optical_design}
\end{figure}

\begin{figure}
  \centering
  \includegraphics[width=1\textwidth]{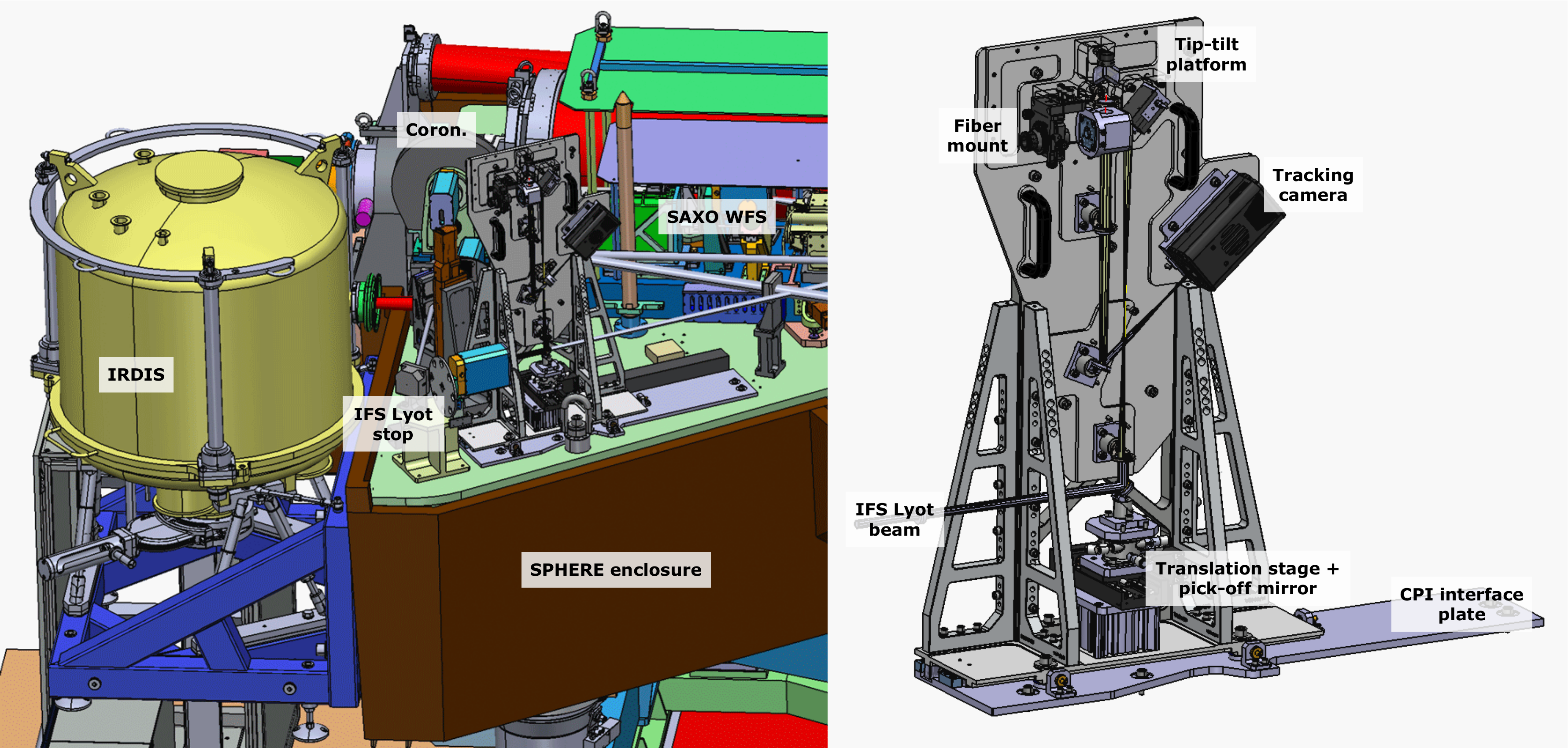}
  \caption{\emph{Left:} Overview of the implementation of the FIM inside of SPHERE. \emph{Right:} Detailed view of the mechanical design of the FIM. The main elements are highlighted on the figure. An additional cover (not shown) will be implemented over the injection optics and fiber mount at the top of the FIM.}
  \label{fig:fim_implementation_full}
\end{figure}

\begin{figure}
  \centering
  \includegraphics[width=1\textwidth]{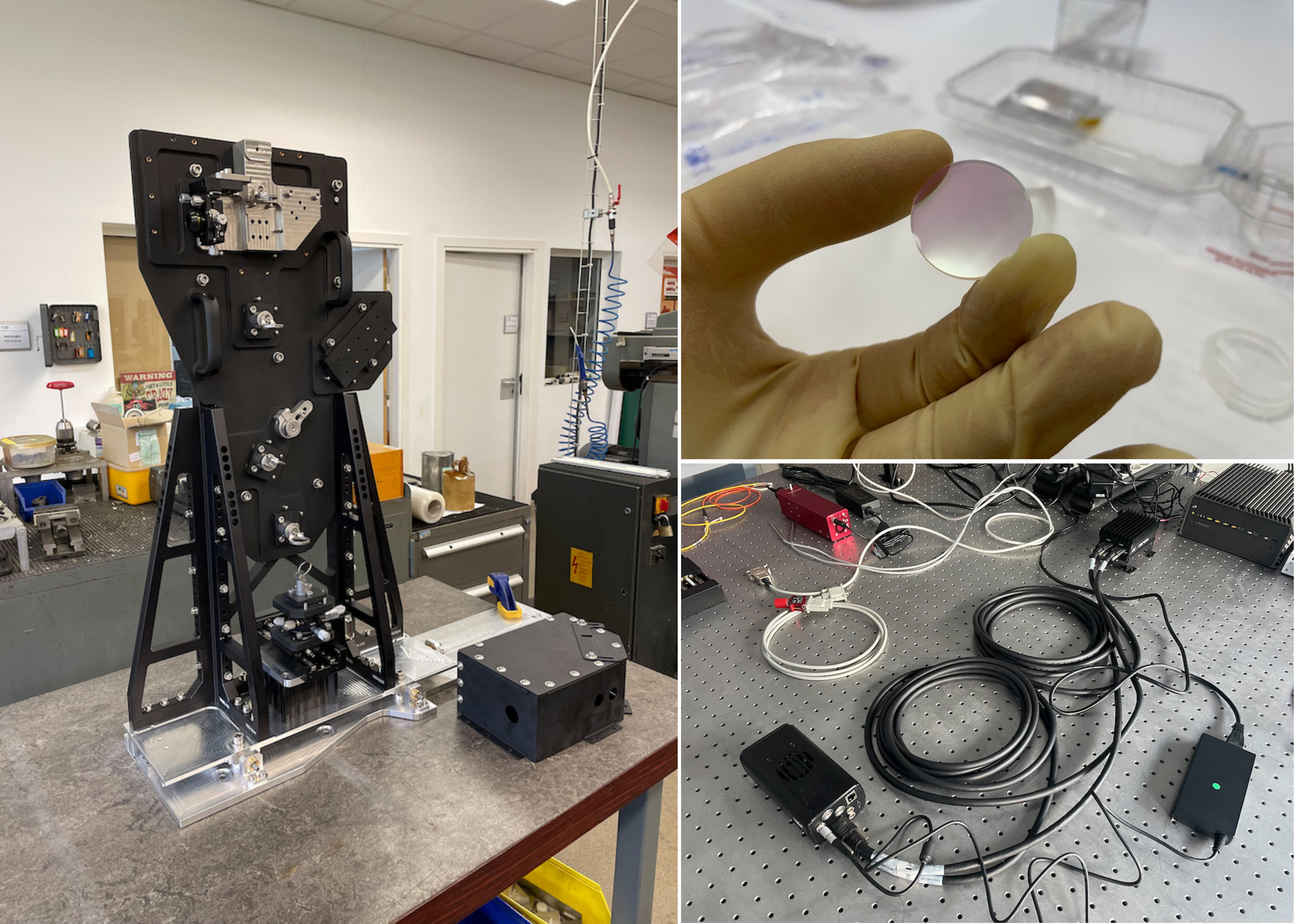}
  \caption{Pictures of some of the final FIM hardware. \emph{Left:} pre-assembly of the FIM mechanical structure. \emph{Top right:} custom dichroic filter. \emph{Bottom right:} FLI C-RED 2 tracking camera and its Pleora frame grabber.}
  \label{fig:fim_pictures}
\end{figure}

The FIM is an additional sub-system implemented inside of the SPHERE enclosure, on the main CPI bench. It diverts the beam of the SPHERE/IFS towards a dedicated optical bench. The optical design with photon sharing is presented in Fig.~\ref{fig:fim_optical_design} and the mechanical implementation is presented in Fig.~\ref{fig:fim_implementation_full}. Images of the final hardware being integrated or tested in the lab are presented in Fig.~\ref{fig:fim_pictures}.

Originally, the VLT/HiRISE system has been proposed for both the $H$ and $K$ bands\cite{Vigan:2018}. However, following the simulation study presented in [\citenum{Otten:2021}], the expected performance in $K$-band was considered to be insufficient to justify the added complexity of implementing the $K$ band in the system. Most of all, the $K$ band requires the use of ZBLAN fibers that provide a high-tranmission at wavelengths longer than 1.6\,\mic, which corresponds to the typical wavelength cutoff of pure silica fibers. However, ZBLAN fibers are expensive and not easy to handle without suffering damages. Moreover, few industrial companies are ready to accept the risk to work with this type of fibers for manufacturing a long fiber bundle such as the one required for HiRISE. For these reasons, we decided to drop support for the $K$ band and focus only on the $H$ band using a pure silica fiber, reference 1310M-HP, manufactured by Coherent (formerly Nufern).

In the FIM, the pick-off mirror is immediately followed by a relay lens that reimages the telescope pupil on a piezo tip-tilt stage. This tip-tilt is used to move the focal-plane image with respect to the fiber bundle and the tracking camera. The beam is (almost) recollimated by a custom cemented doublet, and then a custom dichroic filter designed and manufactured by the Fresnel institute\footnote{\url{https://www.fresnel.fr/}} that splits the wavelengths between the tracking channel (0.95--1.2\,\mic) and the science channel (1.4--1.8\,\mic). Finally, an air-gap doublet is used to reimage the focal-plane at F/3.5 on top of the science fibers in the bundle. F/3.5 does not exactly match the F-number expected by fibers (NA=0.16; F/3.3), but adopting a slightly larger value was mandatory to free some physical space for the mechanical implementation of the system. In the tracking channel, a doublet reimages the focal plane on the tracking camera, a C-RED 2 from First Light Imaging, with an focal ratio of F/40. The images on the tracking camera are Nyquist-sampled starting at 0.95\,\mic.

A retro-injection channel is included in the system to emit light from calibration fibers included in the fiber bundle and reimage them on the tracking camera on top of the astrophysical scene. These calibration fibers will be used for the acquisition procedure to center the planet's PSF on the science fiber, as detailed in [\citenum{ElMorsy:2022}]. Because the beam is not exactly collimated at the level of the dichroic filter, a corrector plate with a very long focal length was added in front of the corner cube to compensate for the small additional length between the dichroic and corner cube. The wavelength of the calibration source is at 1.3\,\mic, i.e. in the transition region of the dichroic filter.

The system is implemented on a vertical bench above the SPHERE/IFS beam. The pick-off mirror is introduced in the beam with an OWIS precision translation stage, which will be controlled from the SPHERE control software. Most mechanical parts are custom made from aluminum or Invar 36, manufactured by the mechanical workshop at the Georg-August-Universität in Göttingen. The other parts are off-the-shelf components from Thorlabs and Newport. The optics are a mix between off-the-shelf lenses and mirrors from Thorlabs and Edmund, and custom optics manufactured by \emph{Optique Fichou}\footnote{\url{https://www.optique-fichou.com/}} (France).

The fiber bundle will be held in a multi-axis fiber optic positioner from Newport using a custom interface piece. The multi-axis positioner will be used for the fine positioning of the bundle with respect to the injection optics. The focus and tip and tilt of the bundle were used as compensator in the tolerancing analysis of the optical design, so these degrees of freedom are essential for a fine alignment of the system.

\subsection{Fiber bundle}
\label{sec:implementation_fb}

\begin{figure}
  \centering
  \includegraphics[width=1\textwidth]{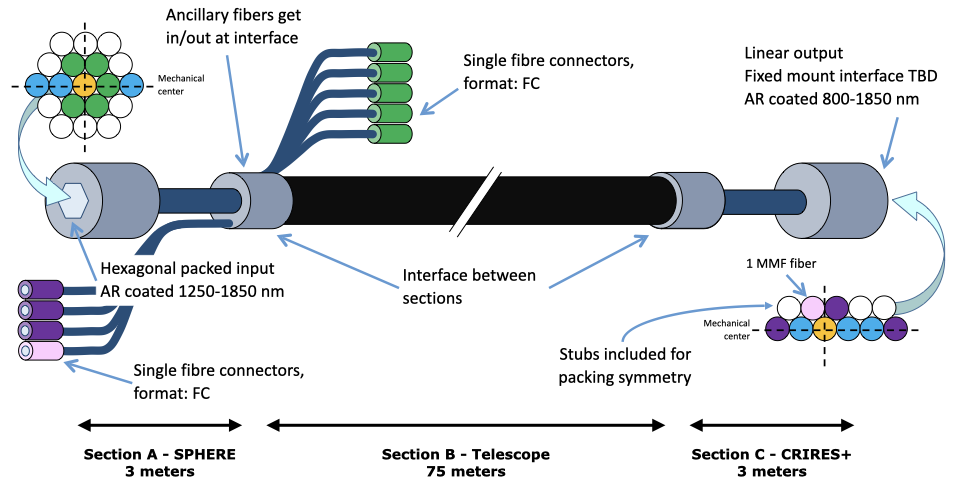}
  \caption{Global concept for the fiber bundle. The bundle is a single piece of hardware with no intermediate connectors in order to maximize the transmission of the astrophysical signal. It is divided in three sections: section A on the SPHERE side, section B around the telescope structure, and section C on the CRIRES+ side. Section B is more heavily protected than sections A and C, which lie in relatively protected environment inside of SPHERE and CRIRES+. The bundle includes 4 science fibers and a number of additional fibers used for target acquisition and for calibration. The bundle will be manufactured by Fibertech Optica (FTO) in Canada.}
  \label{fig:fb_concept}
\end{figure}

A schematic of the proposed design for the fiber bundle is presented in Fig.~\ref{fig:fb_concept}. The bundle is a single piece of hardware composed of three main sections. It features four science fibers (one dedicated to the planet) and ancillary fibers for calibration purposes. The bundle will be manufactured by Fibertech Optica\footnote{\url{https://fibertech-optica.com/}} (FTO) in Canada.

Section A is located on the SPHERE side and has a length of $\sim$3\,m. Most of its length is located inside of the SPHERE enclosure. A part of section A is located outside of the enclosure because the ancillary calibration fibers get in/out of the bundle at the level of the interface between section A and B, and they connect to calibration sources located inside the HiRISE electronics cabinet. The input of the bundle features an hexagonal pattern of fibers that includes the four science fibers, one fiber connected to a fluxmeter, and four retro-injection fibers connected to a calibration source at 1.3\,\mic. The purpose of these five calibration fibers is further described in [\citenum{ElMorsy:2022}]. The input of the bundle will be directly AR-coated with a coating covering both the science and calibration wavelengths.

Section B is the long section of the bundle that goes around the telescope structure. It is more heavily protected because it will be exposed to the harsh environment of the telescope where it can be crushed or stepped on. It includes a central glass rod and a tensile element to maintain the length, minimize the mechanical stresses on the fibers inside, and protect the fibers from breaking. The route around the telescope has been investigated by the Paranal observatory and a steel cable has been pulled all the way to verify the routing feasibility and measure the exact length (75\,m).

Finally, section C is located on the CRIRES+ side and has again a length of $\sim$3\,m. In the output ferrule, the fibers are arranged in a double line. The main line includes the four science fibers plus two external calibration fibers that will be used during AIT; this line of fibers will be aligned along the slit of the spectrograph. The second line includes two calibration fibers, which will be used for the MACAO system to maintain the deformable mirror flat during the HiRISE observations. One of the two calibration fibers is a SMF identical to the science fibers, and the other is a multi-mode fiber (MMF) with a 50\,\mic core. Again, the output of the bundle will be directly AR-coated with a coating covering both the science and calibration wavelengths.

The interface parts between sections A, B and C have been specified to minimize impact during the installation of the bundle at the telescope. For the installation, it is foreseen to start the routing on the SPHERE side, which has both the main bundle and ancillary fibers, and then route section B and C step-by-step around the telescope. It was originally proposed to have connectors at the interfaces between sections to make the installation easier, but designing connectors with losses lower than 0.2\,dB revealed to be extremely risky and costly, and the final level of losses could not be guaranteed. 

\subsection{Fiber extraction module}
\label{sec:implementation_fem}

\begin{figure}[h!]
  \centering
  \includegraphics[width=1\textwidth]{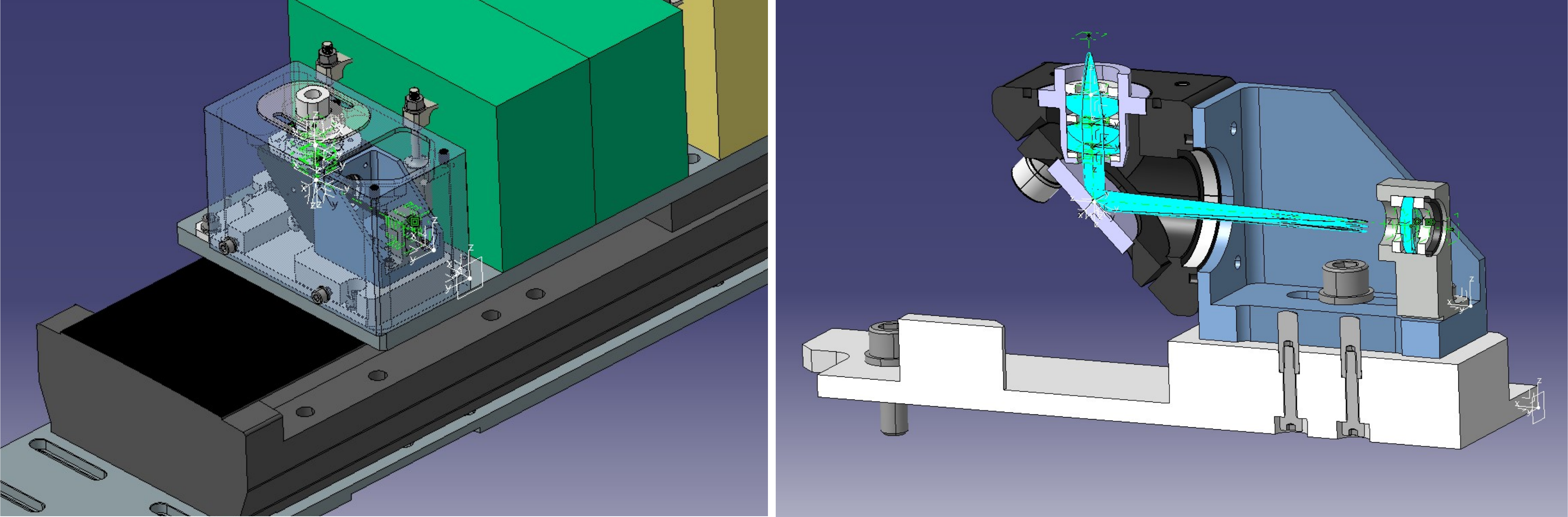}
  \caption{\emph{Left:} Overview of the implementation of the FEM of the CSU at the entrance of CRIRES+. \emph{Right:} Detailed view of the opto-mechanical design of the FEM with a cut through the module. The interface part at the top that will hold the ferrule of the fiber bundle is not the final one in this drawing. The full module is about 15\,cm-long and 8\,cm-wide.}
  \label{fig:fem_implementation}
\end{figure}

The FEM is implemented on the carrier stage unit (CSU) of the warm bench of CRIRES+. It is used to reimage the output of the science and calibration fibers in the VLT focal plane at F/15, which is the focal ratio expected by CRIRES+ for direct observations with the VLT. The opto-mechanical implementation of the FEM on the CSU is presented in Fig.~\ref{fig:fem_implementation}.

Compared to the FIM, the FEM is a much simpler system with no active device to be controlled. Its role is to simply reimage the fibers in the VLT focal plane. To make the installation easier and avoid unnecessary bending of the fibers, a folding mirror is implemented in the FEM so that the bundle's ferrule is simply inserted vertically at the top of the FEM. Since it is installed in the visitor gas cell slot of CRIRES+, the FEM must be easily removable in a matter of minutes. It is held in place by 3 screws that can be accessed without removing the cover. The FEM can then be put back in place quickly using a set of calibrated shims and the precision pins available on the CSU.

The FEM does not include active elements, but it must be able to deal with the MACAO adaptive optics system of CRIRES+, which is located downstream in the optical train. One of the difficulties of the MACAO system is that its deformable mirror cannot be maintained flat in open-loop on the timescale of hours, which are foreseen for the observations with HiRISE. To cope with that limitation, the fiber bundle include two calibration fibers (1 SMF, 1 MMF) located nearby the science fibers, which will be fed by a broadband source and will serve as an AO guide star for the MACAO system. Only one of the two will be used, but to keep some flexibility two different types of fibers were included in the design of the bundle: the SMF will be totally unresolved so that it truly resembles a stellar object, but the transmitted flux might be too low for MACAO; the MMF will be slightly resolved with its large 50\,\mic core, but it will provide a much higher flux. The two options will be tested during the commissioning.

\subsection{Electronics and control}
\label{sec:implementation_electronics}

\begin{figure}
  \centering
  \includegraphics[width=0.8\textwidth]{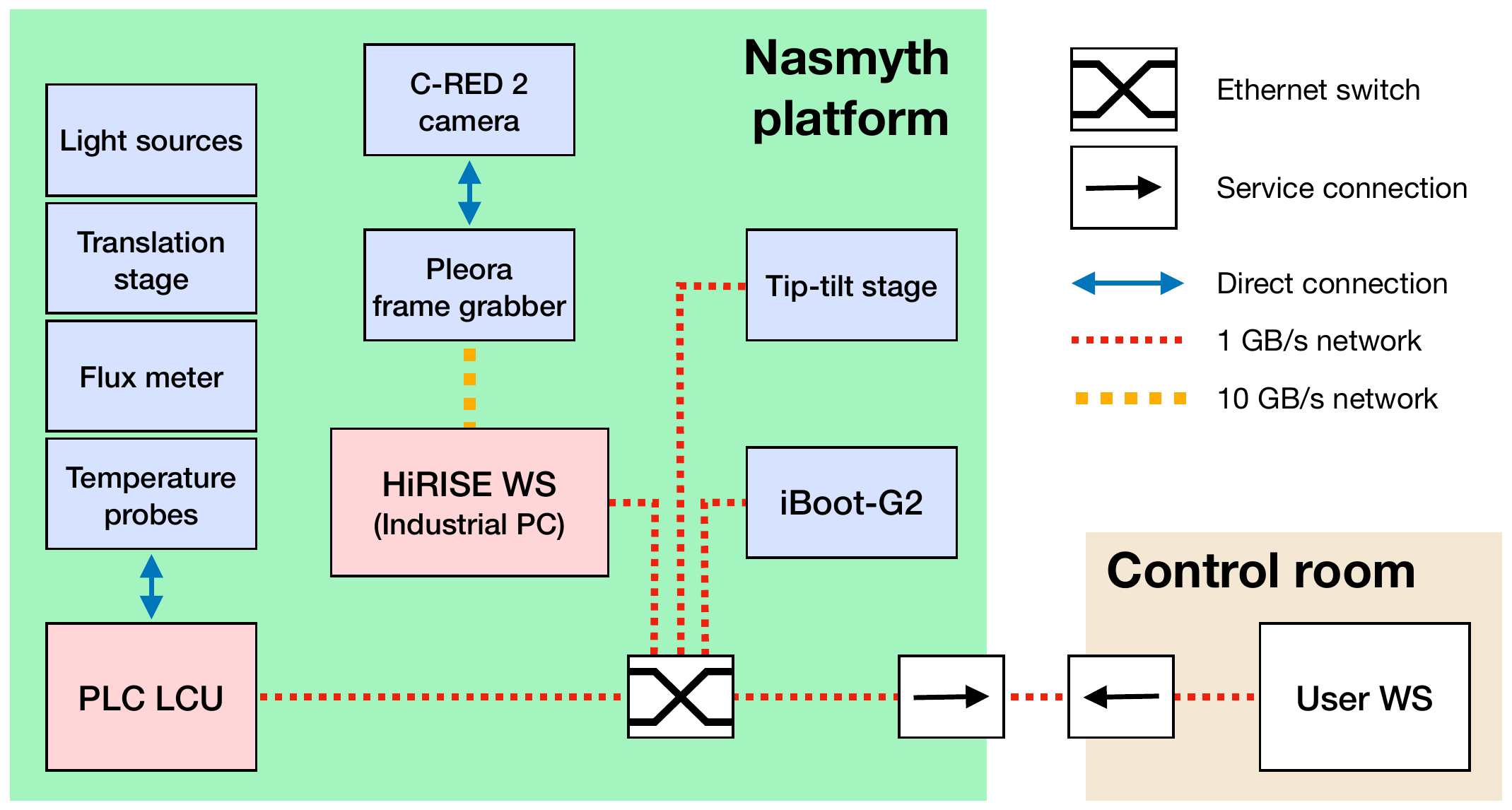}
  \caption{Devices and network implementation of HiRISE. The HiRISE workstation is directly embedded in the HiRISE electronics cabinet to make the installation of the system as easy as possible.}
  \label{fig:network_implementation}
\end{figure}

\begin{figure}
  \centering
  \includegraphics[width=1\textwidth]{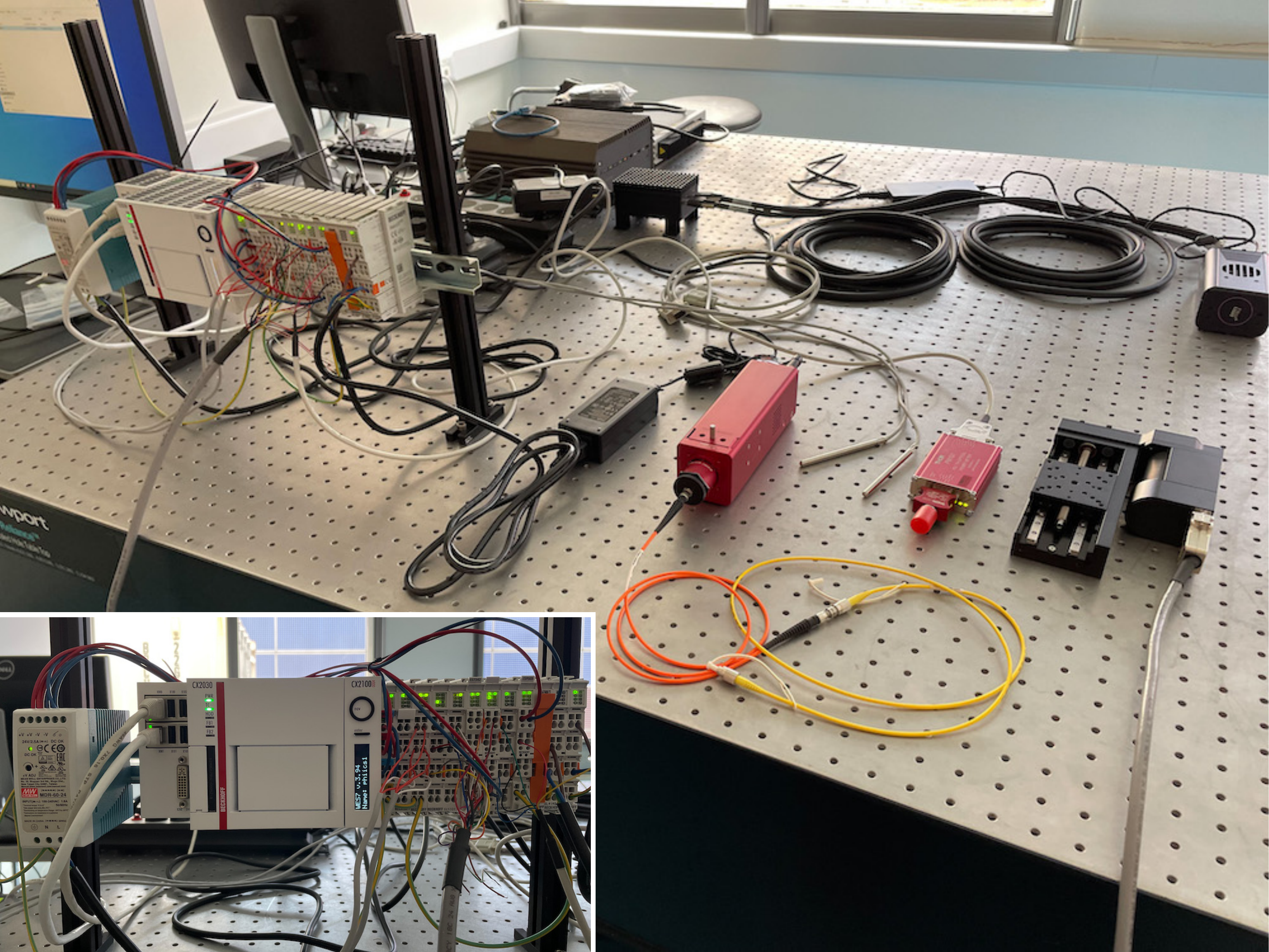}
  \caption{Programmable logic controller (PLC) for the FIM, which controls some of the devices: calibration sources, temperature probes, flux meter and pick-off mirror translation stage. Other important devices such as the tip-tilt mirror and the tracking camera are controlled directly by the workstation through a private network.}
  \label{fig:fim_plc}
\end{figure}

\begin{figure}
  \centering
  \includegraphics[width=0.5\textwidth]{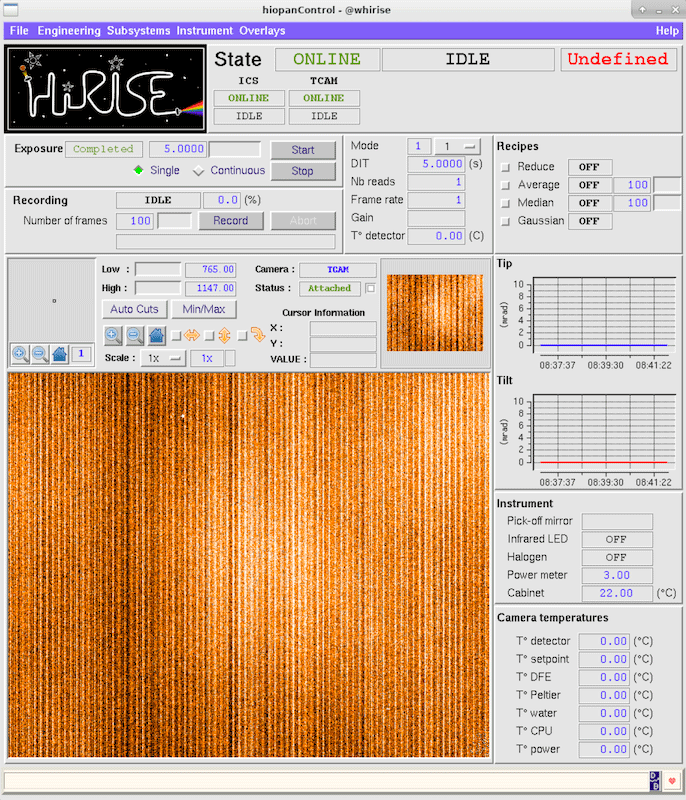}
  \caption{Graphical user interface of the FIM control panel for HiRISE. The HiRISE instrument software is entirely developed following the VLT Software standards.}
  \label{fig:software_gui}
\end{figure}

The FIM in HiRISE includes several active and passive devices, which must be interfaced with the HiRISE workstation for operating the system. All the controllers are located in the HiRISE electronics cabinet (HiCab), which will be installed underneath the SPHERE instrument on the Nasmyth platform. To simplify the implementation, the HiRISE workstation is directly embedded within the HiCab in the form of an industrial computer running VLT Software (CentOS Linux 7.7 + VLT common software 2020). 

The HiRISE workstation is connected to the devices and sensors through different means (Fig.~\ref{fig:network_implementation}). The iPORT frame grabber from Pleora Technologies is connected to the workstation using a dedicated 10GB/s network interface and short-range optical link. The tip-tilt stage controller (E272 from PI) and the connected power switch (iBoot-G2) are connected on the private HiRISE instrument network through a Cisco switch. Finally, the other devices and sensors are connected to a Beckhoff PLC computer, which is itself connected to the private HiRISE network to be accessible from the workstation. Pictures of the hardware in the integration clean room at LAM are presented in Fig.~\ref{fig:fim_plc}.

Finally, a dedicated control software and interface is in development for the control of the system. The main interface is presented in Fig.~\ref{fig:software_gui}. The HiRISE control software follows the same logic and organization as all VLT instruments in order to make the future operations as transparent as possible for the observatory staff. In particular, it is foreseen to develop dedicated templates for the main calibrations of FIM, as well as for the target centering, which corresponds to the accurate placement of the planet's PSF on the science fiber. The target (star) acquisition will be performed by the SPHERE system, and the scientific data acquisition by the CRIRES+ system.

\section{EXPECTED PERFORMANCE}
\label{sec:expected_perf}

\begin{figure}
  \centering
  \includegraphics[width=0.8\textwidth]{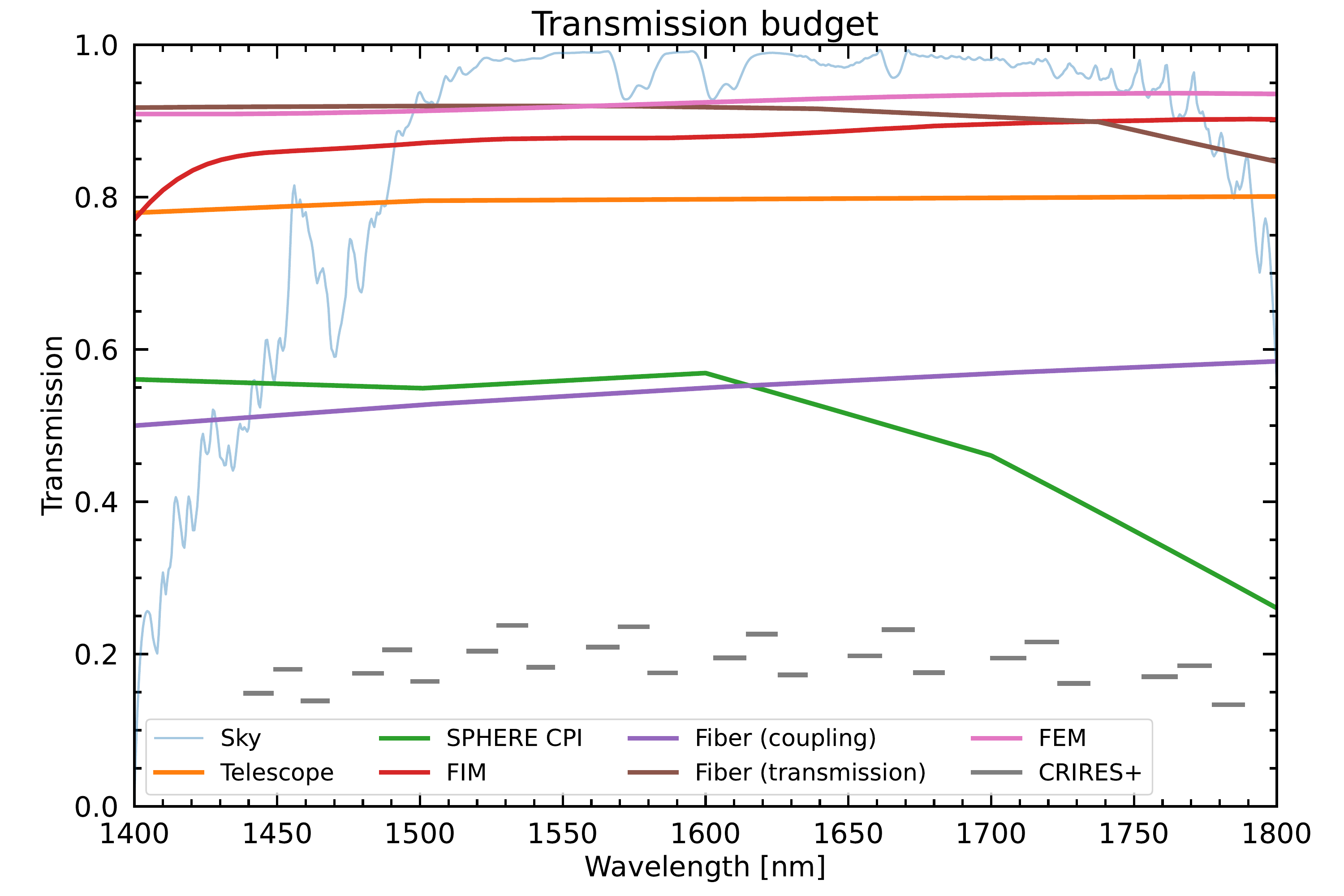}
  \includegraphics[width=0.8\textwidth]{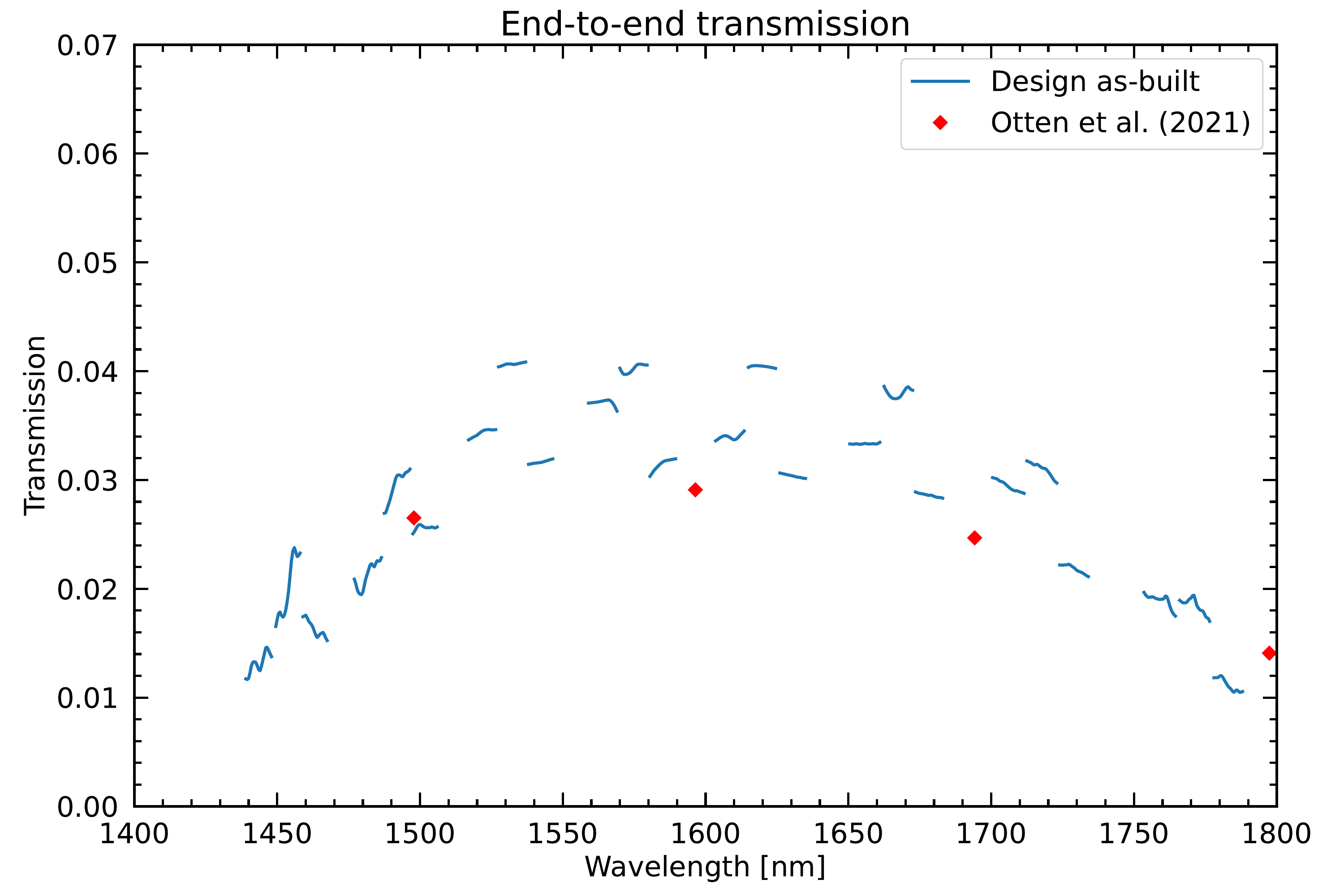}
  \caption{\emph{Top:} Individual components in the transmission budget for HiRISE, based on the final hardware (FIM, FEM), on simulations (fiber coupling), on lab measurements and theoretical values (fiber transmission), on measurements provided by the instrument consortia (SPHERE, CRIRES+), and on theoretical values (sky, telescope). See Sect.~\ref{sec:expected_perf} for more detailed explanations. \emph{Bottom:} Expected end-to-end transmission of the system based on the transmission budget. The plot includes a comparison with the values published in [\citenum{Otten:2021}], which were the current baseline for the development of the project.}
  \label{fig:transmission_budget}
\end{figure}

The end-to-end simulations of [\citenum{Otten:2021}] have clearly demonstrated that the final detection performance of HiRISE is driven by the overall transmission of the system. The transmission budget has been updated based on the final optical components and AR coatings. The complete budget is presented in Fig.~\ref{fig:transmission_budget}.

For SPHERE and CRIRES+ we rely on values that have been measured either in the lab or on sky, which have been kindly provided by the two instrument consortia. For the lenses in the FIM and the FEM, we use the AR coatings information provided by the optics manufacturers. For the mirrors, purchased from Edmund, we use a value 0.98 for the reflectivity at all wavelengths.

One of the important budget contributors is the fiber bundle, and in particular the coupling efficiency of the planet's PSF into the science fiber. For the fused silica fiber itself, we use the attenuation values provided in [\citenum{Cozmuta:2020}] and scaled to the attenuations provided at 1300 and 1500\,nm by Coherent for the 1310M-HP fiber (which are very slightly worse than the values from [\citenum{Cozmuta:2020}]). For the coupling efficiency we use the HiRISE simulation model\cite{Otten:2021} to compute the overlap integrals between the planet's PSF electric field and the field transmitted by the science fiber. For this computation, we include ExAO residuals (for a 0.8\as seeing), SPHERE non-common path aberrations measured with ZELDA\cite{N'Diaye:2016,Vigan:2019,Vigan:2022,pyZELDA:2018}, and the wavefront errors of the FIM based on the Zemax model with as-built optics. We also include a centering error of $\sim$8\,mas ($\sim$0.2\,$\lambda/D$ at 1.6\,\mic) on the science fiber, which corresponds to a realistic pointing error after centering the planet on the science fiber\cite{ElMorsy:2022}.

The updated transmission budget shows a small gain of $\sim$20\% at 1.6\,\mic compared to the values previously reported in [\citenum{Otten:2021}]. This is encouraging for the final performance of HiRISE, which should hopefully be better than anticipated from previous end-to-end simulations. An update of the astrophysical simulations is currently on-going in preparation for the on-sky validation. The gains are mainly due to the high quality of the AR coatings for the custom optics, the quality of the coatings for the FIM dichroic, and the use of highly transmissive fused silica fibers. The current model does not take into account the transmitance of the glasses used for the custom lenses. This may have a slight negative impact that will be evaluated in the near future.

A significant contributor to the transmission budget is the SPHERE \texttt{DICH-K} dichroic, which is used to send the $K$ band light to IRDIS and the $YJH$ bands to the IFS and HiRISE. This dichroic shows a strong decrease of its transmission starting at 1.6\,\mic, which is clearly visible in the top panel of Fig.~\ref{fig:transmission_budget}. As part of the SPHERE+ project\cite{Boccaletti:2020}, an updated version of that dichroic could be considered to improve the transmission budget of HiRISE and the SPHERE IFS. Another possibility to improve the global transmission would be to switch to a dedicated fiber-fed spectrograph such as the VIPA concept\cite{Bourdarot:2018}, which recently demonstrated a transmission of the order of 40\% in H-band on sky Carlotti et al. (this conference).

\section{PROJECT STATUS AND SCHEDULE}
\label{sec:project_status}

HiRISE has followed a step-by-step approach for its validation by ESO. First, the science case for the instrument was submitted to ESO's observing programmes committee (OPC) in September 2019, which expressed a strong scientific support for the project at the end of 2019 and offered the technical aspects to be evaluated by the Paranal observatory and the ESO's scientific and technical committee (STC). Then, a visitor instrument proposal containing a full technical description was submitted to the STC in April 2020, leading to a validation by the STC and ESO's council at the end of 2020 and the acceptance of HiRISE as a visitor instrument for the VLT. The project will now hold a final review of the project with the observatory at the end of July 2022. This final review is not a final design review in the common sense of the term but a final validation of the technical aspects and interfaces that have already been thoroughly discussed between the project and the observatory. As a result, the procurement of the hardware has already started since 2020 (from long lead-time elements) and is now almost complete.

The HiRISE project is therefore entering its main AIT phase in Europe. At the time of writing (end of June 2022), all the hardware for the FIM has been received at LAM, which is the most complex part of the whole system. The bundle is expected to be delivered by FTO in October 2022, and the mechanical parts of the FEM will start to be manufactured over the summer. A ``SPHERE simulator'' is currently being aligned to deliver a beam with the same properties as the SPHERE IFS beam in which the FIM will be placed. In parallel, a fine pre-assembly of the FIM and its optics will be performed using metrology. Then, the SPHERE simulator and FIM will be aligned relative to each other so that the final optical alignment of the FIM can be performed. Because the fiber bundle will arrive later in the year and because the bundle will not easily be usable to assess the optical quality, we anticipate the use of a Phasics near-infrared wavefront sensor in the injection focal plane to evaluate the optical quality of the alignment. The procedure for the validation of the full system including the bundle and the FEM is still under study.

The current schedule foresees an installation of the system at the VLT between February and March 2023. Although this schedule may shift in case of major issues during AIT in Europe, slots have already been allocated in the observing schedule for period P110. These slots will be dedicated to the installation, which requires the telescope to be put offline for some days and nights, and to the on-sky validation of the system to assess the final performance. Assuming a reasonable level of performance, scientific observations could start at the beginning of P111 in April 2023.

HiRISE is a visitor instrument, which means that it will not be offered for observations to the community as a regular facility instrument. However, anyone interested to observe with HiRISE is cordially invited to contact the project to discuss opportunities. The project anticipates the submission of a large observing programme to observe most known sub-stellar companions observable from Paranal but additional programmes in collaboration with the HiRISE project team are a highly desirable possibility.

\acknowledgments

This project has received funding from the European Research Council (ERC) under the European Union’s Horizon 2020 research and innovation programme, grant agreements No. 757561 (HiRISE) and 678777 (ICARUS), from the \emph{Commission Spécialisée Astronomie-Astrophysique} (CSAA) of CNRS/INSU, and from the \emph{Action Spécifique Haute Résolution Angulaire} (ASHRA) of CNRS/INSU co-funded by CNES, and from Région Provence-Alpes-Côte d'Azur under grant agreement 2014-0276 (ASOREX).

\bibliography{paper}
\bibliographystyle{spiebib}

\end{document}